\documentclass[journal]{IEEEtran}
\usepackage{ifpdf}

\ifCLASSOPTIONcompsoc
  \usepackage[nocompress]{cite}
\else
  \usepackage{cite}
\fi

\ifCLASSINFOpdf
  \usepackage[pdftex]{graphicx}
\else
  \usepackage[dvips]{graphicx}
\fi

\usepackage{amsmath}
\usepackage[linesnumbered,ruled]{algorithm2e}
\usepackage{array}
\usepackage{mdwmath}
\usepackage{mdwtab}
\usepackage{multirow}
\usepackage[caption=false,font=footnotesize,labelfont=sf,textfont=sf]{subfig}

\usepackage{url}
\usepackage{listings}

\begin{document}

\title{ArcText: A Unified Text Approach to Describing Convolutional Neural Network Architectures}

\author{Yanan~Sun,~\IEEEmembership{Member,~IEEE,}
	    ~Ziyao~Ren,
		~Gary~G.~Yen,~\IEEEmembership{Fellow,~IEEE,} 
        ~Bing~Xue,~\IEEEmembership{Member,~IEEE,}
        ~Mengjie~Zhang,~\IEEEmembership{Fellow,~IEEE,}
        and Jiancheng Lv,~\IEEEmembership{Member,~IEEE,}

\thanks{Yanan Sun, Ziyao Ren and Jiancheng Lv are with the College of Computer Science, Sichuan University, Chengdu 610065, China (e-mails: ysun@scu.edu.cn; ziyaoren99@gmail.com, lvjiancheng@scu.edu.cn).}
\thanks{Gary G. Yen is with the School of Electrical and Computer Engineering, Oklahoma State University, Stillwater, OK 74078 USA (e-mail:gyen@okstate.edu).}
\thanks{ Bing Xue, and Mengjie Zhang are with the School of Engineering and Computer Science, Victoria University of Wellington, PO Box 600, Wellington 6140, New Zealand (e-mails: bing.xue@ecs.vuw.ac.nz; and mengjie.zhang@ecs.vuw.ac.nz).}

}% <-this % stops a space

\IEEEtitleabstractindextext{%
\begin{abstract}
The superiority of Convolutional Neural Networks (CNNs) largely relies on their architectures that are often manually crafted with extensive human expertise. Unfortunately, such kind of domain knowledge is not necessarily owned by each of the users interested. Data mining on existing CNN can discover useful patterns and fundamental sub-comments from their architectures, providing researchers with strong prior knowledge to design proper CNN architectures when they have no expertise in CNNs. There have been various state-of-the-art data mining algorithms at hand, while there is only rare work that has been done for the mining. One of the main reasons is the gap between CNN architectures and data mining algorithms. Specifically, the current CNN architecture descriptions cannot be exactly vectorized to the input of data mining algorithms. In this paper, we propose a unified approach, named ArcText, to describing CNN architectures based on text. Particularly, four different units and an ordering method have been elaborately designed in ArcText, to uniquely describe the same architecture with sufficient information. Also, the resulted description can be exactly converted back to the corresponding CNN architecture. ArcText bridges the gap between CNN architectures and data mining researchers, and has the potentiality to be utilized to wider scenarios.
\end{abstract}

}

\maketitle

\IEEEdisplaynontitleabstractindextext

\IEEEpeerreviewmaketitle

\section{Introduction}
\label{section_introduction}

\IEEEPARstart{D}{iverse} Convolutional Neural Network (CNN)-based models, such as GoogleNet~\cite{szegedy2015going}, ResNet~\cite{he2016deep}, DenseNet~\cite{huang2017densely}, to name a few, have been purposely designed for addressing different machine learning tasks, and demonstrated their superior performance largely in the field of computer vision. Their well-designed architectures have been principally recognized as the key component contributing to the superiority. Specifically, GoogleNet presented the parallel-node architecture collectively feeding their outputs as the input to the sole consequent node; ResNet introduced the architecture with addition-based skip connections architecture from one node to its adjacent node; and DenseNet developed the architecture having concatenation-based skip connections to another node that receives the output of all its previous nodes.

Designing an optimal CNN architecture is often of high-cost: manually starting from a skeleton architecture and then refining it based on feedback from a number of trial-and-error tests until the satisfactory performance is reached. This procedure highly depends on human expertise in CNNs and domain knowledge of the task at hand, which are used to provide guidelines to the refinement~\cite{sun2019evolving}. However, both requirements are not necessarily held by the end-users, resulting in limited application scenarios. In addition, the CNN architecture well-designed on one task generally cannot be directly reused when the task changes, and the manual procedure needs to perform once again. Recently, the research of Neural Architecture Search (NAS)~\cite{elsken2018neural} has been raised, aiming at reducing the human expertise intervention as much as possible during the design of CNN architectures.

NAS formulates the architecture design as an optimization problem that is often discrete, highly constrained, with multiple conflicting objectives and computationally expensive~\cite{sun2019completely}. Evolutionary algorithms~\cite{back1996evolutionary} and reinforcement learning~\cite{sutton1998reinforcement} are the dominating optimizers of solving NAS because of their promising characteristics in effectively addressing complex optimization problems. Although the NAS algorithms, to some extent, have reduced the requirements of human expertise and domain knowledge in designing CNN architecture, most of them suffer from relying on extensive computational resource~\cite{sun2019surrogate}. For example, the reinforcement learning-based NAS method proposed in~\cite{zoph2016neural} consumed 28 days using 800 Graphics Processing Units (GPUs), while the large-scale evolution NAS method~\cite{real2017large} employed 250 GPUs over 11 days. In principle, the intensive computational resource-dependent problem is caused by training CNNs from scratch that is computationally expensive. The training time of one CNN often varies from several hours to dozens of days on one GPU card even for median-scale datasets, such as CIFAR10~\cite{krizhevsky2009learning}. In NAS, a number of CNNs will be trained using the same training routines, thus demanding a large amount of computational resource for performing the algorithms. Unfortunately, sufficient computational resource for sustaining the running of NAS algorithms is not necessarily available to end-users. As a result, the low-cost CNN architecture design is highly desired but still remains a challenging issue.

\begin{figure}[htp]
	\centering
	\includegraphics[width=0.9\columnwidth]{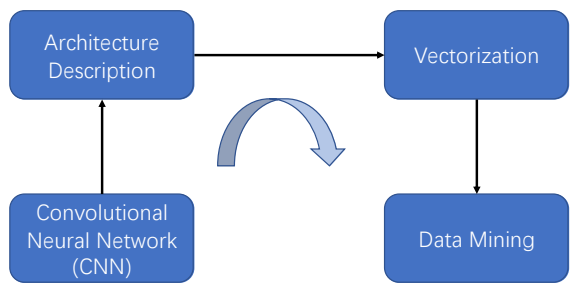}\\
	\caption{An example to illustrate the flow chart of data mining for CNN architectures Firstly, the CNN model is provided for the mining. Then, its architecture is described with the corresponding methods. Next, the description is vectorized to the form that the computer programs can accept. At last, the vectorization result is inputted into the data mining algorithms.}\label{fig_vectorization}
\end{figure}

Data mining of existing CNN architectures potentially provides an alternative to the low-cost CNN architecture design. First, crucial components of CNN architectures could be discovered for a group of similar tasks via the mining, which would provide a strong prior knowledge to CNN architecture design, and consequently promote the manual design even when the users are with poor knowledge~\cite{tej2018determining}. Second, mining the relationship between the architectures and their performance could help to build an effective and efficient regression model that could be used to replace the computationally expensive training process during NAS, and naturally addresses the intensive computational resource-dependent problems of the existing NAS algorithms~\cite{sun2019surrogate}. Last but not least, numerous CNN architectures are being manually designed because they can effectively solve challenging machine learning tasks, which made the data mining of CNN architectures practicable in terms of the available data volume~\cite{szegedy2015going,he2016deep,huang2017densely}. To the best of our knowledge, only very few works of mining CNN architectures have been reported publicly. The major bottleneck is believed caused by the current CNN architecture description methods. Particularly, data mining algorithms receive the numerical values as its input, while the description of CNN architectures cannot be exactly transformed to the numerical values that are fed to the data mining algorithms. Note that this transformation is also called \textit{vectorization}. For the convenience of the understanding, the flow chart of data mining on a CNN architecture is shown in Fig.~\ref{fig_vectorization}. Specifically, given the CNN model to be mined, its architecture is firstly described with the description methods, and then the description result is transformed into the numerical data, through the vectorization, which can be regarded as the input data to the data mining algorithms, for the further mining. This process can be easily understood through the face recognization function of iPhone to automatically unlock either mobile phone or cell phone. During the face recognization, the camera of the iPhone captures the face image of the person first, and then the pixel values of the face image are fed to the recognization algorithm embedded into the iPhone to judge whether or not the face is the owner of the iPhone. In this example, the person, the face image, the pixels values of the image, and the recognization algorithm can be viewed as the CNN, the description of the CNN architecture, the vectorization result, and the data mining algorithm, respectively. 

Existing methods for describing CNN architectures can be generally classified into three different categories based on common practice. They are the image-based description methods, the Natural Language (NL)-based description methods, and the hybrid description methods. Specifically, the image-based methods employ images to describe CNN architectures via visualizing the architecture framework, while some details, such as the kernel sizes, number of feature maps, which are very important to the performance of CNN, cannot be revealed due to the limited layout. Directly using the pixel values of images is a common way to vectorize images with all features. Obviously, this vectorization method cannot be used for image-based description due to the loss of the important CNN architecture information. 

The NL-based methods utilize text of NL to describe CNN architectures. Compared to the image-based methods, this method can provide all the details of CNN architectures. NL Processing (NLP) techniques have provided multiple commonly used algorithms to vectorize text based on the grammar rules. However, they cannot be used for the text describing CNN architectures. This is because the NLP techniques are mainly designed for the word text recording the events related to people and the words have a steady grammar rule that can be easily recognized by the human at different levels. Unfortunately, there is no any grammar rule to describe CNN architectures. In this situation, different researchers may give significantly different text descriptions to CNN architectures, and even the same researcher may generate different text description for the same CNN architecture under different occasions. The hybrid methods, on the other hand, are based on certain hybridizations of both methods mentioned above, where the image is used to illustrate the framework of the CNN architecture while the NL is used to compensate for the details that cannot be shown on the image. However, the vectorization for the hybrid method is still challenging due to the use of NL. Moreover, the vectorization for the hybrid method also involves the feature extraction from cross-domain data, which is still an infant research area and there is no effective solution available. 

In this paper, we propose a text approach to describing CNN architectures, denoted as \textbf{ArcText}, to bridge the data mining algorithms and CNN architectures, by addressing the aforementioned limitations from the existing description methods. The contributions of the proposed ArcText method are summarized below:

\begin{enumerate}
    \item ArcText can generate unique text description for a given CNN architecture, and the generated descriptions can also be exactly translated back to the CNN architecture. Specifically, four units are elaborately designed for handling the nodes of CNN architectures. Furthermore, a method has been designed to uniquely order the nodes in CNN. This establishes the foundation to mine CNN architectures.

\item ArcText can describe almost all existing CNN architectures, including the state-of-the-art ones manually designed and those that can be generated by existing NAS algorithms. This provides a convenient way to efficiently store and exchange CNN architectures. Upon this, CNN researchers can easily share their CNN architectures and the corresponding information, which will provide sufficient data samples needed for data mining algorithms on CNN architectures.

\item ArcText is based on text. Thus, the advanced NLP techniques can be easily applied based on the descriptions generated by ArcText, which provide an economical way to mine CNN architectures using the NLP algorithms at hand.
\end{enumerate}

The remainder of the paper is organized as follows. Firstly, related works on describing CNN architectures are reviewed and commented, and the findings justify the necessity of the proposed ArcText method in Section~\ref{section_literature}. Next, the details of ArcText are documented in Section~\ref{section_algorithm}. Then, two representative examples are illustrated in Section~\ref{sec_example} to help readers intuitively realize ArcText. Finally, the conclusions and future work are provided in Section~\ref{section_conclusion}.

\begin{figure*}[htp]
	\centering
	\includegraphics[width=1.9\columnwidth]{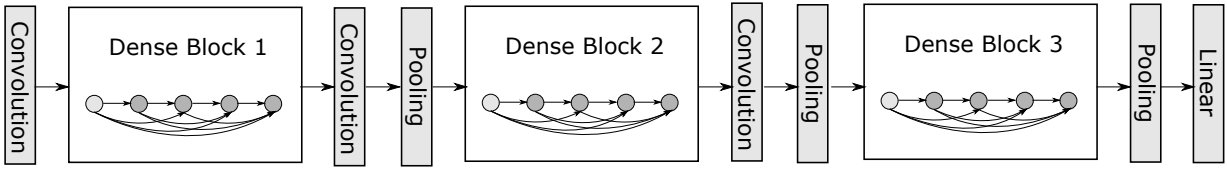}\\
	\caption{An example of the DenseNet architecture described using the image. Note that this image is the same as that shown in its seminal paper~\cite{huang2017densely}.}\label{fig_densenet}
\end{figure*}

\section{Related Work}
\label{section_literature}
In this section, the existing methods on describing CNN architectures are reviewed, and then the proposed algorithm is justified in terms of its necessity. 

\subsection{Natural Language (NL)-Based Methods}
\label{sec2_1}
The NL-based methods are mainly used to talk about CNNs among CNN researchers. A CNN is often deep, varying from dozens to thousands of nodes, therefore, different people may adopt completely different ways to describe the CNN architectures using the NL-based methods. For example, given a CNN, some people may describe the overall architecture first, and then supplement the details; while other people may directly start to describe the details node by node, or describe the architectures based on a well-known CNN first, and then provide the supplemental detailed information. Although this method could result in different descriptions of CNN architectures, it does not harm human's understanding owing to the powerful functions of human brains. However, when the resulted description is directly inputted to the data mining algorithms, they are not able to understand because the computer programs have a different way of understanding text. This is different from the NLP domain, where there have been multiple state-of-the-art algorithms to convert the languages into the values that data mining algorithms can process as their inputs. The reason is that the NLP targets at the languages used in daily communication, by following a basic grammar rule no matter whether the NL is English, Chinese or other languages. As of now, there is no such grammar rule in place for CNN architectures.

\subsection{Image-Based Methods}
\label{sec2_2}
The image-based methods work well in demonstrating the overview of CNN architectures. Because it is an intuitive way to help the understanding of CNNs, some deep learning libraries have provided the corresponding toolkit to generate such images, such as the TensorBoard from TensorFlow~\cite{AbadiTensorFlow}. The TensorBoard can automatically generate the image of the corresponding CNN architecture when CNN is implemented by TensorFlow. The major limitation of this method is that the images of CNN architectures can only show their brief information concerning mainly on the topology, such as how many nodes in the CNN and how the nodes are connected. The other information, such as the configurations of convolutional layers, pooling layers, and fully-connected layers cannot be displayed because these images cannot properly show too many details easily. Although using the pixel values of images is a common way to vectorize images, the requirements are based on that the image has included sufficient features of the object. Due to the missing configurations of CNN architectures from the images, the pixels of CNN images cannot be used.

An example of DenseNet~\cite{huang2017densely} is shown in Fig.~\ref{fig_densenet} to conveniently discuss the limitation of such kinds of methods for describing CNN architectures using images. Particularly, this example is the same as that provided by its author in \cite{huang2017densely}. As shown in this example, it is convenient for us to understand the architecture of DenseNet, especially, the significant difference from other existing state-of-the-art CNNs, i.e., the dense blocks. However, if we use the pixel values of this image as the input to the data mining algorithms, e.g., another neural network-based data mining algorithm, it is clear that we cannot obtain the exact results because the details of the dense blocks are now shown in this image. In addition, the configurations of the convolutional layer, pooling layers and the fully-connected layers are not shown in this image. In principle, any CNNs like DenseNet can use this image to demonstrate their architectures. However, these CNNs may have different performance and their own distinctive configurations.

\subsection{Hybrid Methods}
\label{sec2_3}
The hybrid methods are achieved by using NL and images to collectively describe CNN architectures, which is the method most commonly used, and many state-of-the-art works employ such a way to demonstrate their architectures~\cite{he2016identity,huang2016deep} in their seminal publications. This is because the images could provide an overview of the CNN architecture, while the NL-based text can complement the details that cannot be shown on the images. However, due to the problem suffered from the NL-based methods mentioned in Subsection~\ref{sec2_1}, the description resulted by the hybrid method still cannot be used directly for mining the CNN architectures. In addition, extracting CNN architecture information from images and NL-based text is a cross-domain research problem~\cite{elkahky2015multi}, which is very challenging to the current data mining community.

\subsection{Necessity of the Proposed Method}

The conclusion can be drawn from the discussions on Subsections~\ref{sec2_1} to \ref{sec2_3} that the current image-based methods cannot describe the complete information of CNN architectures. Consequently, the corresponding images cannot be used for mining CNN architectures. Although the missing information can be additionally provided by developing advanced image-based methods, several new problems will inadvertently come along. Firstly, it is difficult to assign complete information for generating the same pixel values for the same CNN architecture. For example, the configurations of the CNN kernels are the key components of CNN architectures, and any blank space in the images can be used to assign such kinds of configurations. As a result, each assignment will result in unique pixel values for the same CNN architecture. Secondly, if the whole information of the CNN architecture renders in the image, the essence of visualization techniques will be lost. This is because the visualization techniques aim at intuitively showing the information. With too many details showing in the image, the resulted image will become not intuitive. The hybrid methods are also not suitable because of their partial use of images for the description.

Although the NL-based methods cannot be used as discussed above, they provide the potential to address the limitations of the above methods, by supplementing some rules to describe CNN architectures. Some recent works have taken the first step, such as the Peephole method~\cite{deng2017peephole} and the E2EPP method~\cite{sun2019surrogate}. Specifically, Peephole and E2EPP methods proposed the text-based description methods to describe CNN architectures for mining the relationship between CNN architectures and their respective performance. However, both can only be used to describe the CNN architectures that are generated by their own NAS algorithms, and cannot be adapted to most of the existing CNN architectures. Another aspect motivated from the NL-based methods is that there have been many state-of-the-art NLP algorithms~\cite{manning1999foundations,collobert2008unified,mou2016convolutional}, which can be conveniently used to mine CNN architectures. In addition, there are also some promising algorithms for vectorizing text, such as the word2vec~\cite{rong2014word2vec,goldberg2014word2vec}. However, such algorithms are designed based on the corpus. With the proposed ArcText method, collecting the sufficient corpus of CNN architectures will become easy, and some existing vectorizing algorithms can be readily used to promote the data mining of CNN architectures.

\section{The Proposed ArcText Algorithm}
\label{section_algorithm}
The proposed ArcText method aims at describing CNN architectures in a unified way based on text, so that the result can show a unique description of the corresponding CNN architecture, building the foundation to fully mine CNN architectures. 

\subsection{Algorithm Overview}
The design of ArcText is motivated by the natural language that people used for communications, where words are the basic units of sentences, and a sentence typically describes an independent event by combining these words. A CNN is composed of multiple nodes which can be classified into two categories: building blocks containing convolutional layers, pooling layers and fully-connected layers, and operations including the activation function, Batch Normalization (BN)~\cite{NIPS2017_6790}, etc. The design of ArcText is to provide a grammar rule to describe the nodes of a CNN, and a unified way to combine these nodes for the description sentence.

In ArcText, we design three different types of units to describe three types of building blocks, and an additional unit to describe the operations. For each of the units, we design a group of properties to distinguishes layers that are with the same type of building blocks, and the same operations with different configurations. For the convenience of describing ArcText, we collectively call the four units as ArcUnits. By changing the property values, the ArcUnit enables itself the ability to distinguish the building blocks and operations with different configurations, thus forming a unique text-based description for the CNN.
The properties of an ArcUnit can be classified into three different types. The first contains only the property of identifier indicating the position of the corresponding node in the CNN. The second is composed of the basic properties referring to configurations of the corresponding node. The third consists of the auxiliary property concerning the connection information of the corresponding node. When all nodes of the CNN have been described, the ArcUnits are connected with an increasing order of the identifier values to form the complete description of the CNN architecture.

\begin{figure*}[htp]
	\centering
	\includegraphics[width=1.9\columnwidth]{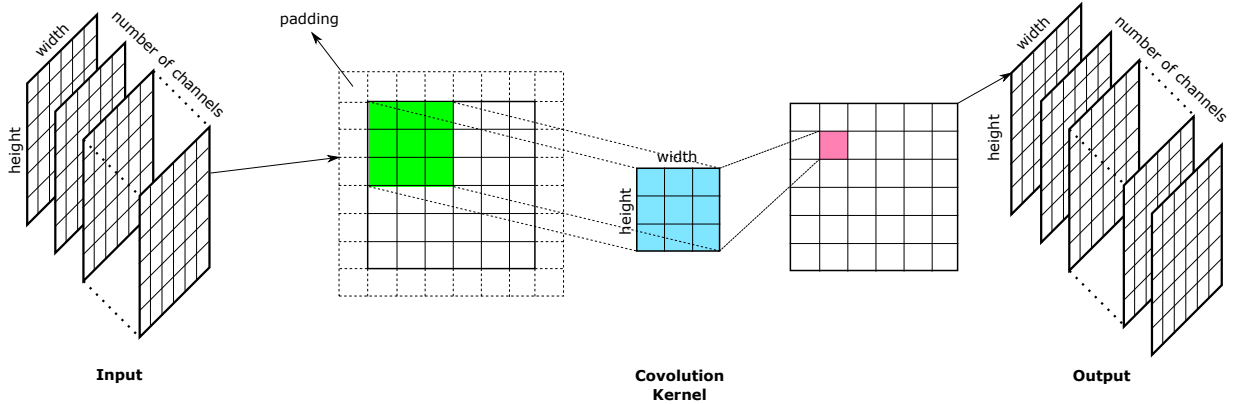}\\
	\caption{An illustrated example of the convolutional operation which is composed of three parts: the input data, the convolutional kernels, and the output data.}\label{fig_conv_example}
\end{figure*}

\begin{algorithm}
	\label{alg_framework}
	\caption{Framework of ArcText}
	\KwIn{The CNN $C$ for description.}
	\KwOut{The description of $C$.}
	$ArcUnits\leftarrow$$\emptyset$\;
	\label{alg_gramework_step1_start}
	\For{each node $l$ in $C$}
	{
		$u\leftarrow$ Choose a proper ArcUnit based on the type of $l$\;
		Set the values of the basic properties of $u$\;
		$ArcUnits\leftarrow$ $ArcUnits\cup u$\;
	}	\label{alg_gramework_step1_end}
	
	\For{each unit $u$ in $ArcUnits$}
	{\label{alg_gramework_step2_start}
		$l\leftarrow$ Find the corresponding node of $u$ in $C$\;
		$i\leftarrow$ Locate the position of $l$ in $C$\;
		Assign $i$ as the identifier value of $u$\;
	}\label{alg_gramework_step2_end}
	Describe the auxiliary property of each unit in $ArcUnits$\;\label{alg_gramework_step3}
	Combine the units in $ArcUnits$ based on their identifier values  in ascending order\;\label{alg_gramework_step4}
	\textbf{Return} the combination.
\end{algorithm}
The proposed ArcText method is mainly composed of four steps as shown in Algorithm~\ref{alg_framework}. First, the information of each node in the given CNN $C$ is used to set the basic property values of the corresponding ArcUnits (lines~\ref{alg_gramework_step1_start}-\ref{alg_gramework_step1_end}). Then, the position of each node in $C$ is located, and used to set the identifier value of the corresponding ArcUnit (lines~\ref{alg_gramework_step2_start}-\ref{alg_gramework_step2_end}). Next, the auxiliary property values of the ArcUnits are specified (line~\ref{alg_gramework_step3}), which is based on the identifier values. Finally, the description of $C$ is generated by combining the ArcUnits with ascending order of their identifier values (line~\ref{alg_gramework_step4}). Note that a CNN must be provided in advance as the input to perform ArcText. Furthermore, the provided CNN does not need to follow a particular format, but only sufficient information needs to be used. Here, the ``sufficient information'' means that the provided CNN can be manually implemented for successfully running on computers. In the following subsections, the details of the four steps are documented individually.

\subsection{Set Basic Property Values of ArcUnits}
\label{sec3_1}
As have mentioned above, the proposed ArcText method provides four different types of ArcUnits, and each ArcUnit has three types of properties. Particularly, the four types of ArcUnits are ConvArcUnit, PoolArcUnit, FullArcUnit, and Multi-Function Unit (MFUnit), which are used to describe convolutional layers, pooling layers, fully-connected layers, and operations, respectively\footnote{Note that, although some recently NAS algorithms~\cite{sun2019completely,sun2018automatically} claimed that CNNs can still achieve the state-of-the-art performance without using the fully-connected layers, we still consider the fully-connected layers in the proposed ArcText method for the compatibility and generality of previous CNNs.}. Specifically, the ConvArcUnit, PoolArcUnit, and FullArcUnit are designed mainly by considering the properties of convolutional layers, pooling layers, and fully-connection layers, respectively, and the MFUnit is designed by considering those that the three above units cannot cover. In addition, their designs also consider their topologies within CNN. By setting different values to the properties of ArcUnits, the nodes belonging to the same type can be differentiated. In the following, the properties of four units are introduced, based on which the details of setting the property values are documented.

\subsubsection{\textbf{ConvArcUnit}} 
\label{sub_section_conv}
The basic properties of ConvArcUnit are motivated by the convolutional operation. For the convenience of the development, an example of the convolutional operation is shown in Fig.~\ref{fig_conv_example}. Specifically, the convolutional operation takes effect on the input data (the left part in this example), through the convolutional kernels (the central part in this example), and finally generates the corresponding output (the right part in this example). The input data and the output data commonly have three dimensional, i.e., the width, the height, and the number of the channels. The convolutional kernel is commonly a two-dimensional matrix with some learnable weights (the two dimensions are often called as the height and the width of this convolutional kernel). During the operation, the convolutional kernel travels through the input data with a predefined stride having the step sizes at the horizontal and vertical directions. In some situations, the output is required to be kept to a desirable size. To achieve this, the extra values are needed to be padded to the four directions of the channel data. This process is called as the ``padding''. Commonly, one channel of the input data has one corresponding convolutional kernel. However, some state-of-the-art CNNs, such as the GoogleNet, have demonstrated promising performance of using multiple channels of the input sharing the same convolutional kernels. This process is called as ``groups'' which can benefit to lower the number of the learnable parameters for improving the generalization ability~\cite{cohen2016group,ioannou2017deep}. In most CNN operations, the convolutional kernel and its mirror in the input data have the same sizes. Recently, the ``dilation'' has been designed to improve the receptive field of CNNs, to promote the corresponding performance~\cite{li2018csrnet,wolterink2016dilated,yu2017dilated}, where the size of the mirror is enlarged with a predefined parameter between each element.

\begin{table}[!htp]
	\renewcommand{\arraystretch}{1.2}
	\caption{The properties of ConvArcUnit for describing convolutional layers.} \label{table_convarcunit_info}
	\begin{tabular}{|m{0.03\columnwidth}|m{0.15\columnwidth}|m{0.65\columnwidth}|}
		\hline
		\multicolumn{1}{|c|}{\textbf{No.}} & \multicolumn{1}{c|}{\textbf{Name}} & \multicolumn{1}{c|}{\textbf{Remark}} \\
		\hline
		1 & id & the identifier with an integer value to denote the position of the associated convolutional layer in the network\\
		\hline
		2 & in\_size & a three-element tuple with integer values to denote the width, the height, and the number of channels of input data\\
		\hline
		3 & out\_size & a three-element tuple with integer values to denote the width, the height, and the number of channels of output data\\
		\hline
		4 & kernel & a two-element tuple with integer values to denote the width and the height of convolutional kernels\\
		\hline
		5 & stride & a two-element tuple with integer values to denote the vertical and the horizontal steps when moving the kernels \\
		\hline
		6 & padding & a four-element tuple with integer values to denote the padding information at up, down, left and right directions, respectively, each element is composed of two sub-elements denoting the value and number of the padding operation at the direction\\
		\hline
		7 &dilation & an integer to denote the space size between kernel elements for the convolutional operation \\
		\hline 
		8 & groups & an integer to denote the number of input channels that use the same feature map\\
		\hline
		9 & bias\_used & a boolean number to indicate whether the bias term is used or not\\
		\hline
		10 & connect\_to & a tuple consisting of the identifiers to which it connects\\
		\hline
	\end{tabular}
\end{table}

\begin{figure*}[htp]
	\centering
	\includegraphics[width=1.9\columnwidth]{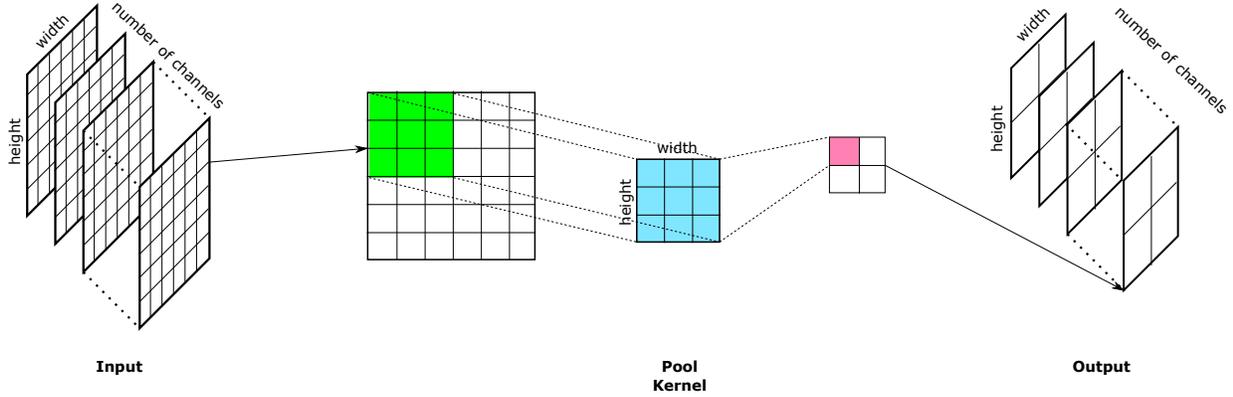}\\
	\caption{An illustrated example of the pooling operation which is composed of three parts: the input data, the kernels for the pooling, and the output data.}\label{fig_pool_example}
\end{figure*}

In summary, the ConvArcUnit has the basic properties containing the input size, the output size, the kernel size, the stride size of the kernel, the number of padding, the padding mode, the spacing size between kernel elements, the number of the input channels using the same feature map, and the property indicating whether or not the bias term is used. The auxiliary property contains the collections of identifiers in which it will connect to. The details of the property of the ConvArcUnit are shown in Table~\ref{table_convarcunit_info}, where the first column denotes the number of the properties, the second column refers to the property names, and the third column lists the remarks of the properties. Note that we have merged the number of padding and the padding mode into one property (the 6-th property shown in Table~\ref{table_convarcunit_info}) for ConvArcUnits for the reason of simplicity. As can be seen from Table~\ref{table_convarcunit_info}, a ConvArcUnit has 10 different properties. Specifically, the properties of $id$, $in\_size$, $out\_size$, $kernel$, $stride$, $padding$, $dilation$, $groups$ use the integer values. The $bias\_used$ adopts the boolean value to represent its status of enabled or disabled. The $conect\_to$ is a tuple containing the identifiers of the units where this unit will connect.

\subsubsection{\textbf{PoolArcUnit}} The PoolArcUnit is designed based on the pooling operation which is illustrated in Fig.~\ref{fig_pool_example}. As can be seen from this figure and Fig.~\ref{fig_conv_example}, both the pooling operations and the convolutional operations are very similar, except that the pooling operation has no ``padding''. In addition, the pooling operation also does not have the property of ``groups'' which aims to reduce the number of learnable parameters. That is because the pooling operation has no learnable parameters. Furthermore, in the pooling operation, the input and the output have the same number of the channels, which is not always true for the convolutional operation. Overall, a PoolArcUnit has the basic properties of the input size, the output size, the kernel size, the stride size, the number of padding, the spacing size between the kernel elements, and the number of the input channels using the same kernel. Because there are two types of pooling operation, i.e., the max pooling and the average pooling, another basic property is also designed to denote whether it is a max pooling layer or an average pooling layer. Compared with the ConvArcUnit, the PoolArcUnit does not have the property representing the padding mode although it has the property indicating the number of padding. The reason is that the pooling operation employs zeros for the padding by default. Furthermore, utilizing other values for the padding is not valid for pooling operation. In addition, the ArcPoolUnit employs the same auxiliary properties as those of the ArcConvUnit.

\begin{table}[!htp]
	\renewcommand{\arraystretch}{1.2}
	\caption{The properties of PoolArcUnit for describing pooling layers.} \label{table_poolarcunit_info}
	\begin{tabular}{|m{0.03\columnwidth}|m{0.15\columnwidth}|m{0.65\columnwidth}|}
		\hline
		\multicolumn{1}{|c|}{\textbf{No.}} & \multicolumn{1}{c|}{\textbf{Name}} & \multicolumn{1}{c|}{\textbf{Remark}} \\
		\hline
		1 & id & the identifier with an integer value to denote the position of the associated pooling layer in the network\\
		\hline
		2& type & a string denoting the type of pooling layer\\
		\hline
		3 & in\_size & a three-element tuple with integer values to denote the width, the height, and the number of channels of input data\\
		\hline
		4 & out\_size & a three-element tuple with integer values to denote the width, the height, and the number of channels of output data\\
		\hline
		5 & kernel & a two-element tuple with integer values to denote the width and the height of pooling kernels\\
		\hline
		6 & stride& a two-element tuple with integer values to denote the vertical and the horizontal steps when moving the kernels \\
		\hline
		7 & padding & a four-element tuple with integer values to denote the padding information at up, down, left and right directions, respectively\\
		\hline
		8 & dilation & an integer to denote the space size between kernel elements for the pooling operation \\
		\hline
		9 & bias\_used & a boolean number to indicate whether the bias term is used or not\\
		\hline
		10 & connect\_to & a tuple consisting of the identifiers to which it connects\\
		\hline
	\end{tabular}
\end{table}

Table~\ref{table_poolarcunit_info} shows the details of the properties for the PoolArcUnit, where the first, the second and the third columns denote the numbers, the names and the remarks of the properties, respectively. The PoolArcUnit has 10 properties, all of which employ the same value types as those of the ConvArcUnit, in addition to the $type$ that is chosen from ``Avg'' and ``Max'' referring to the average pooling operation and the max pooling operation, respectively.

\subsubsection{\textbf{FullArcUnit}}
The FullArcUnit is designed based on the fully-connected layers. Compared to the ConvArcUnit and the PoolArcUnit, the FullArcUnit has fewer properties. Particularly, we have designated the basic properties of $in\_size$, $out\_size$ to denote the input size and the output size, respectively, and they employ the same data types as those of the ConvArcUnit and the PoolArcUnit. In addition, the FullArcUnits also employ the same auxiliary properties as those of the ConvArcUnits and PoolArcUnits. Table~\ref{table_fullarcunit_info} lists the details of the properties for the FullArcUnit.

\begin{table}[!htp]
	\renewcommand{\arraystretch}{1.2}
	\caption{The properties of FullArcUnit for describing fully-connected layers.} \label{table_fullarcunit_info}
	\begin{tabular}{|m{0.03\columnwidth}|m{0.15\columnwidth}|m{0.65\columnwidth}|}
		\hline
		\multicolumn{1}{|c|}{\textbf{No.}} & \multicolumn{1}{c|}{\textbf{Name}} & \multicolumn{1}{c|}{\textbf{Remark}} \\
		\hline
		1 & id & the identifier with an integer value to denote the position of the associated fully-connected layer in the network\\
		\hline
		2 & in\_size & an integer value to denote the size of the input data\\
		\hline
		3 & out\_size & an integer value to denote the size of the output data\\
		\hline
		4 & connect\_to & a tuple consisting of the identifiers to which it connects\\
		\hline
		
	\end{tabular}
\end{table}

\subsubsection{\textbf{MFUnit}}
The MFUnit is designed to describe the operations in CNNs, such as the activation function, BN, Dropout~\cite{srivastava2014dropout}, etc. In practice, these operations are often utilized in conjunction with the convolutional layers and fully-connected layers. However, they are not designed as part of the properties of ConvArcUnits and FullArcUnits. In the following, we will explain the reasons in detail.

First, if they are designed as the properties of the corresponding ArcUnits, the use of the ArcText will become complex and not flexible as well. For example, if the BN operation is available to the ConvUnits, another property indicating whether or not the BN operation is applied should also be designed accordingly. That is because CNNs which were popular 10 years ago did not use the BN operation. In the meanwhile, if the property denoting the activation function is designed for the ConvUnits, another property indicating the order of convolutional operation, BN and this activation operation should also be designed. That is because some state-of-the-art CNNs adopted the order of ``convolutional operation$\rightarrow$BN$\rightarrow$activation function’’, while others used the order of ``convolutional operation$\rightarrow$activation function$\rightarrow$BN’’. Clearly, because of the design of the properties in relate to BN and activation function, additional properties must also be designed for the adaption, which consequently increase the complexity of using them. 

Secondly, if they are designed for the ArcUnits, the proposed ArcText method cannot describe some existing state-of-the-art CNNs. For example, if the BN operation is designed to the ConvArcUnit, the BN is used only when the convolutional layers appear. However, the large-scale evolutional method, which is a state-of-the-art NAS method, has proven a promising CNN architecture that a BN can be used independently to the use of the convolutional layer. In addition, for the multiple data with different shapes collectively serving as the input to the same unit, which is very popular among the state of the arts, their shapes must be transformed to the same size. This transformation is often realized by using the same number of convolutional layers as that of the multiple data. In this case, if these operations, i.e., the activation function, BN, etc., are designed together with the convolutional layers, another operations will be performed in conjunction with the convolutional operation, which is not necessary.

\begin{table}[!htp]
	\renewcommand{\arraystretch}{1.2}
	\caption{The properties of MFUnit for describing operations.} \label{table_MFunit_info}
	\begin{tabular}{|m{0.03\columnwidth}|m{0.15\columnwidth}|m{0.65\columnwidth}|}
		\hline
		\multicolumn{1}{|c|}{\textbf{No.}} & \multicolumn{1}{c|}{\textbf{Name}} & \multicolumn{1}{c|}{\textbf{Remark}} \\
		\hline
		1 & id & the identifier with an integer value to denote the position of the associated fully-connected layer in the network\\
		\hline
		2 & name & an string value to denote the type of the operation\\
		\hline
		3 & in\_size & an integer value to denote the size of the input data\\
		\hline
		4 & out\_size & an integer value to denote the size of the output data\\
		\hline
		5 & value & a tuple consisting of the necessary parameters of the operation\\
		\hline
		6 & connect\_to & a tuple consisting of the identifiers to which it connects\\
		\hline
	\end{tabular}
\end{table}

The basic properties designed for MFUnits are $name$, $in\_size$, $out\_size$, and $value$, which denote the name of the operation, the input size, the output size, and the value of the operation, respectively. As the designs for the other three ArcUnits, the MFUnit also has the auxiliary property of $connect\_to$. Table~\ref{table_MFunit_info} lists the details of the properties of the MFUnits. Specifically, the $name$ uses a string to denote the operation name, which should be titled based on the conventions. For some operations which have no conventions for their names, we have provided options for their names, such as ``Dropout’’ for the Dropout operation, ``BN’’ for the Batch Normalization operation, ``Interpolation’’ for the interpolation operation, and ``Addition’’ and ``Concatenation’’ for the skip connections like those of ResNet~\cite{he2016deep} and DenseNet~\cite{huang2017densely}, respectively. Note that the $value$ should be a concatenation in alphabet order with string format if the operation has multiple values. The design of MFUnits provides the flexibility and simplicity for the use of ArcText.

\subsubsection{Details of Setting Basic Property Values}
Setting the values of the basic properties of ArcUnits is quite straightforward, i.e., just copying the values of the nodes in the CNN to the corresponding properties of the ConvArcUnits, PoolArcUnits, FullArcUnits, and MFUnits. The reason for not specifying the values of identifiers and the auxiliary property at this stage is that most state-of-the-art CNNs are often graph-like. If we do not have a well-designed method to traverse the CNN, the position of the nodes in the CNN will be changed when describing it at the different occasions. As a result, the resulted description may be different for the same CNN, and thus cannot be used for the mining algorithms, which is inconsistent with the goal of the proposed algorithm. Furthermore, the setting of basic property values is mainly for specifying the identifier values, and the details will be discussed in Subsection~\ref{sec3_2}.

\subsection{Assign Identifier Values}
\label{sec3_2}
As discussed above that the identifier values are the positions of the corresponding nodes in the CNN, thus, the first step of assigning the identifier values is to find the positions of the nodes, which is not an easy task because many state-of-the-art CNN architectures are graph-like, instead of the linear structure where the nodes are connected as a list (i.e., each node of the CNN has only one input and one output, except that the input node has no input). If there is no well-designed algorithm for finding the position of each layer in a CNN, the CNN may have different descriptions that will still suffer from the limitations of the NL-based description method as discussed in Subsection~\ref{sec2_1}. The proposed method for finding the node positions can address this problem, and the details are shown in Algorithm~\ref{alg_position_location}.
\begin{algorithm}
	\label{alg_position_location}
	\caption{Find the Position of Each Node}
	\KwIn{The nodes $L=\{l_1,l_2,\cdots,l_n\}$ of the CNN.}
	\KwOut{$L=\{l_1,l_2,\cdots,l_n\}$ associed with its resective position number.}
	$G\leftarrow$ Construct a directed acyclic graph based on the connections of the nodes in $L$\;
	\label{alg_loc_step1}
	$S\leftarrow$ Find the node of which the indegree and outdegree are $0$ and $1$, respectively, from $G$\;
	\label{alg_loc_step2_start}
	$E\leftarrow$ Find the node of which the indegree and outdegree are $1$ and $0$, respectively, from $G$\;
	Set $1$ and $n$ as the positions of the nodes associated to $S$ and $E$, respectively\;
	\label{alg_loc_step2_end}
	$i\leftarrow 2$\;
	\label{alg_loc_step3_start}
	\While{P($S$, $E$) and E(P($S$, $E$))}
	{
		$path\leftarrow$ Find the longgest path $P_l(S, E)$  and $E(P_l(S, E))$ \;\label{alg_loc_find_longest_path}
		\If{$|path| > 1$}
		{\label{alg_loc_hash_start}
			Descript the nodes of each path using ArcUnits and get their hash values\;
			$path\leftarrow$ Pick the element of which the hash value is the largest\;
			\label{alg_loc_hash_end}
			\If{$|path| > 1$}
			{\label{alg_loc_random_start}
				$path\leftarrow$ Randomly pick one from $path$\;
				\label{alg_loc_random_step}
			}\label{alg_loc_random_end}
		}
		\ForEach{node in $path$}
		{\label{alg_loc_only_one_start}
			Set $i$ as the position of the node associated $node$ if it has not been numbered yet\;
			$i\leftarrow i + 1$\;
		}\label{alg_loc_only_one_end}
	}\label{alg_loc_step3_end}
	\textbf{Return} $L=\{l_1,l_2,\cdots,l_n\}$.
\end{algorithm}

Particularly, finding the positions of $n$ nodes in the CNN is composed of three steps. The first is to construct a directed acyclic graph based on the nodes’ connection (line~\ref{alg_loc_step1}), i.e., there will be an edge from node $a$ to node $b$ if $a$ has a connection pointing to $b$ in the CNN. Note that this is a manual step by reading the information of the provided CNN. The second is to mark the first node (denoted as $S$) and the last node (denoted as $E$) by calculating the in-degree and outdegree of each node in the graph. Clearly, the input node only has the outdegree of one, and the last node only has the indegree of one. Consequently, the positions of both are numbered as $1$ and $n$, respectively (lines~\ref{alg_loc_step2_start}-\ref{alg_loc_step2_end}). The third is to find the  positions of the other nodes (lines~\ref{alg_loc_step3_start}-\ref{alg_loc_step3_end}), which is achieved by finding the longest path from $S$ to $E$ (line~\ref{alg_loc_find_longest_path}) with the condition that this path has at least one node unnumbered, until both cannot be connected by a path having no unnumbered node. If there exists the longest path, just assigning the position of each node according to its order in the path (lines~\ref{alg_loc_only_one_start}-\ref{alg_loc_only_one_end}); otherwise, the hash values of each path will be calculated based on their descriptions generated by ArcUnits, and the one having the largest hash value is viewed as the longest path (lines~\ref{alg_loc_hash_start}-\ref{alg_loc_hash_end}). If there are still multiple paths having the same hash values, a random one is picked up (lines~\ref{alg_loc_random_start}-\ref{alg_loc_random_end}). Note that in this step, $P(S, E)$ denotes that there is a path from $S$ to $E$, $E(P(S, E))$ refers to there exist at least one node unnumbered in $P(S, E)$, and $|\cdot|$ is a cardinality operator. In the following, we will provide the details of calculating the hash value and the reason for doing so.

As shown in Algorithm~\ref{alg_position_location}, the hash values are calculated based on the ArcUnits describing the corresponding nodes in the path. Particularly, the nodes are described one by one based on its order in the path, and then their respective ArcUnits are connected as a string. After that, the hash value of the string is calculated. Note that any hash method can be used here as long as the conflicting problem can be avoided. However, to keep the consistency between different users, we recommend the 224-hash code~\cite{housley2004224} because its implementation is widely available in almost all programming languages and it has no conflicting problems in most application scenarios. Before generating the string through the combination, each ArcUnit is transformed to a short string by connecting its property name and the corresponding values based on the property number using the symbol of ``;''. If there are multiple values for the property, these values are connected with a predefined symbol of ``-''. For example, the kernel size and the stride size of a pooling layer are $(2,2)$ and $(1,1)$, respectively, the string of both property-value pairs is ``kernel:2-2;stride:1-1''. As mentioned above, finding the position of each node in the CNN is to provide the identifier values of the ArcUnits, which could generate the unique description for a CNN. If multiple different descriptions are generated for the same CNN, the corresponding description method clearly cannot be used for data mining algorithms of CNN architectures. Based on the hash values, it can be guaranteed that CNN can be represented by the only one description.

\begin{figure}[!htp]
	\centering
	\includegraphics[width=0.85\columnwidth]{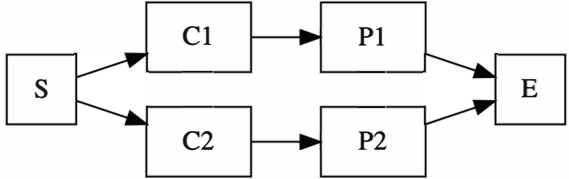}\\
	\caption{A CNN has two branches with the same information, where ``S'' and ``E'' denote the first node and the last node of this CNN, C1 and C2 are the two convolutional layers with the same configuration, and P1 and P2 are the two pooling layers with the same configuration.}\label{fig_same_architecture}
\end{figure}

Note that in Algorithm~\ref{alg_position_location}, there is a random operation as shown in line~\ref{alg_loc_random_step}, which does not change the unique nature of the description. The reason is that the nodes on each path having completely the same information, i.e., the nodes at the same position are the same types, and their configurations are also the same. Thus, no matter which one among them is selected, the resulted description will be the same. To illustrate this situation, an example of the CNN architecture is provided in Fig.~\ref{fig_same_architecture}, where ``S'' and ``E'' denote the input node and the output node, ``C1'', ``C2'', ``P1'', and ``P2'' refer to the two convolutional layers and the two pooling layers, respectively. In addition, ``C1'' and ``C2'' have the same information and result in the same descriptions generated by the ConvArcUnit, which is the same for those of ``P1'' and ``P2'' in addition to the descriptions generated by the PoolArcUnit. Obviously, the path of ``S-C1-P1-E'' has the same description as those of path ``S-C2-P2-E''. As a result, the random operation will give the same description to the CNN, i.e., the description of ``S-C1-P1-E'' or `S-C2-P2-E'' are the same. When the nodes’ position has been confirmed, the identifier values of the ArcUnit will be set based on their corresponding layers.

\subsection{Set Auxiliary Property Values and Combine the ArcUnits}
\label{sec3_3}
Based on the design of the ArcUnits, the auxiliary property is about the connection information, which is represented by the positions of the corresponding nodes. Because the positions have been set as the identifier values of the ArcUnits, the setting of auxiliary property values is just to follow the connections of the corresponding nodes, by copying the corresponding identifier values.

During the stage of combining the ArcUnits, each ArcUnit is transformed to a string based on the details provided in Subsection~\ref{sec3_2} for calculating the hash values first, and then all the ArcUnits are combined based on their identifier values in increasing order. In the proposed ArcText method, the symbol of ``\textbackslash n'' (newline) is used to combine the strings of each ArcUnit for the readability on the text-based description.

\section{Illustrative Examples}
\label{sec_example}
In order to help the readers better understand how the proposed ArcText method works, we present the step-by-step introduction to ArcText for two representative CNNs in this section. Particularly, the first representative is part of the GoogleNet~\cite{szegedy2015going} which obtained the best scores among the classification tasks and the detection tasks of the Large Scale Visual Recognition Challenge (ILSVRC) in 2014. The reason for using it is that this example contains minimal architecture while sufficient information to demonstrate how ArcText works for different cases owing to its completely tree-like architecture. This will be helpful to demonstrate how the proposed ArcText will be used on complex CNN architectures. The other representative is the ResNet~\cite{he2016deep} which was proposed in 2015 and also the champions of classification tasks of ILSVRC in 2015. The ResNet is very popular in practice, and most researchers from the deep learning community are very familiar with it. Thus, the example of ResNet will be more helpful in understanding how the proposed ArcText works. 

In demonstrating each example, we will firstly present the architectures of the representative using the hybrid methods, i.e., using images to show the overview architectures, and then providing the detailed configurations based on text, which is also the method in their respective seminal paper to demonstrate the architecture. After that, we will follow the three major steps in Algorithm~\ref{alg_position_location} to illustrate the details of describing the CNNs. i.e., constructing the graph, finding the path, and then followed by numbering the layer position.

\subsection{Example for the First Representative}
\label{sec_first_example}
\begin{figure}[!htp]
	\centering
	\includegraphics[width=1\columnwidth]{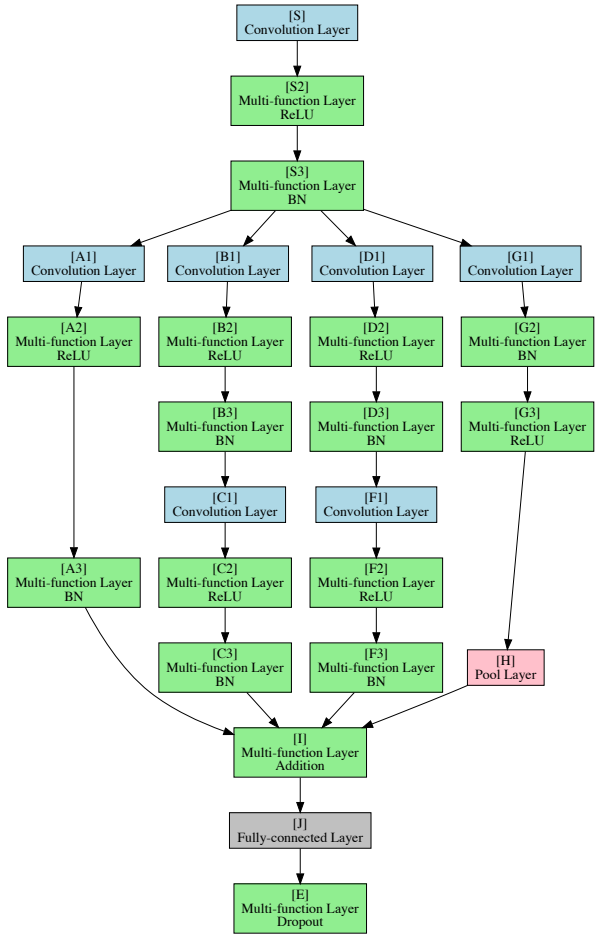}\\
	\caption{The architecture information of the first example CNN. In this figure, each rectangle denotes a node of the CNN. The word inside each block shows the name and the information of the node.}\label{fig_example}
\end{figure}
\begin{table}[htp]
	\renewcommand{\arraystretch}{1.2}
	\caption{The information of the convolutional layers of the first example (Part I).} \label{table_conv_example}
	\centering
	\begin{tabular}{|c|c|c|c|c|c|c|} 
		\hline
		Id&in\_size&out\_size&kernel&stride&padding&dilation \\
		\hline
		S&32,32,3&32,32,3&1,1&1,1&0&1\\
		\hline
		A1&32,32,3&16,16,10&2,2&2,2&0&1\\
		\hline
		B1&32,32,3&32,32,10&1,1&1,1&0&1\\
		\hline
		D1&32,32,3&16,16,3&2,2&2,2&0&1\\
		\hline
		G1&32,32,3&31,31,10&2,2&1,1&0&1\\
		\hline
		C1&32,32,10&16,16,10&2,2&2,2&0&1\\
		\hline
		F1&16,16,3&16,16,10&1,1&1,1&0&1\\
		\hline
	\end{tabular}
\end{table}

\begin{table}[htp]
	\renewcommand{\arraystretch}{1.2}
	\caption{The information of the convolutional layers of the first example (Part II).} \label{table_conv_example_p2}
	\centering
	\begin{tabular}{|c|c|c|c|} 
		\hline
		Id&groups&bias\_used&connect\_to \\
		\hline
		S&1&No&S2\\
		\hline
		A1&1&No&A2\\
		\hline
		B1&1&No&B2\\
		\hline
		D1&1&No&D2\\
		\hline
		G1&1&No&G2\\
		\hline
		C1&1&No&C2\\
		\hline
		F1&1&No&F2\\
		\hline
	\end{tabular}
\end{table}

\begin{table}[!htp]
	\renewcommand{\arraystretch}{1.2}
	\caption{The information of the pooling layer of the first example (Part I).} \label{table_pool_example}
	\centering
	\begin{tabular}{|c|c|c|c|c|c|} 
		\hline
		Id&type&in\_size&out\_size&kernel&stride \\
		\hline
		H&Max&31,31,10&16,16,10&2,2&2,2\\
		\hline
	\end{tabular}
\end{table}

\begin{table}[!htp]
	\renewcommand{\arraystretch}{1.2}
	\caption{The information of the pooling layer of the first example (Part II).} \label{table_pool_example_p2}
	\centering
	\begin{tabular}{|c|c|c|c|c|} 
		\hline
		Id&padding&dilation&bias\_used&connect\_to \\
		\hline
		H&1,0,1,0&1&No&I\\
		\hline
	\end{tabular}
\end{table}

\begin{table}[!htp]
	\renewcommand{\arraystretch}{1.2}	
	\caption{The information of the fully-connected layer of the first example.} \label{table_full_example}
	\centering
	\begin{tabular}{|c|c|c|c|c|} 
		\hline
		Id&in\_size&out\_size&act\_fun&connect\_to\\
		\hline
		J&2560&512&ReLU&E\\
		\hline
	\end{tabular}
\end{table}

\begin{table}[!htp]
	\renewcommand{\arraystretch}{1.2}	
	\caption{The information of the other nodes of the first example.} \label{table_mf_example}
	\centering
	\begin{tabular}{|c|c|c|c|c|c|} 
		\hline
		Id&Name&in\_size&out\_size&value&connect\_to\\
		\hline
		S2&ReLU&32,32,3&32,32,3&Null&S3\\
		\hline
		S3&BN&32,32,3&32,32,3&Null&A1,B1,D1,G1\\
		\hline
		A2&ReLU&16,16,10&16,16,10&Null&A3\\
		\hline
		A3&BN&16,16,10&16,16,10&Null&I\\
		\hline
		B2&ReLU&32,32,10&32,32,10&Null&B3\\
		\hline
		B3&BN&32,32,10&32,32,10&Null&C1\\
		\hline
		C2&ReLU&16,16,10&16,16,10&Null&C3\\
		\hline
		C3&BN&16,16,10&16,16,10&Null&I\\
		\hline
		D2&ReLU&16,16,3&16,16,3&Null&D3\\
		\hline
		D3&BN&16,16,3&16,16,3&Null&F1\\
		\hline
		F2&ReLU&16,16,10&16,16,10&Null&F3\\
		\hline
		F3&BN&16,16,10&16,16,10&Null&I\\
		\hline
		G2&BN&31,31,10&31,31,10&Null&G3\\
		\hline
		G3&ReLU&31,31,10&31,31,10&Null&H\\
		\hline
		I&Addition&16,16,10&16,16,10&Null&J\\
		\hline
		E&Dropout&512&512&0.5&Null\\
		\hline
	\end{tabular}
\end{table}

The topology of the provided CNN example is shown in Fig.~\ref{fig_example}, where each rectangle denotes a node in the CNN. For the convenience of the viewing, we have highlighted the rectangles in the same color if they are the same node types, and also have written their names and information inside the rectangles. Note that the names are provided just for the convenience of understanding how the proposed ArcText works, while in practice, the nodes have no constant name. In addition, their information is provided in Tables~\ref{table_conv_example} to \ref{table_mf_example} which is with the formats of the properties proposed in the four units, except the first columns of these tables that show the names (denoted by ``Id’’) of the corresponding nodes, and the values of ``padding'' column in Table~\ref{table_conv_example} is just a short representation for the real values by using ``0'' to denote ``(0,0),(0,0),(0,0),(0,0)'' defined in Table~\ref{table_convarcunit_info}.

\subsubsection{Constructing Graph} Because the graph construction is straightforward based on the connection information shown in Fig.~\ref{fig_example}, the details of the construction will not be presented here. In addition, the input layer and the output layer have already named as ``S'' and ``E'', respectively. 

\subsubsection{Finding Path}  Based on the provided information, we could obtain the four paths that are ``S-S2-S3-B1-B2-B3-C1-C2-C3-I-J-E'', ``S-S2-S3-D1-D2-D3-F1-F2-F3-I-J-E'', ``S-S2-S3-G1-G2-G3-H-I-J-E'' and ``S-S2-S3-A1-A2-A3-I-J-E'', among which the first two have the same longest lengths. To this end, we set the basic property values of all nodes shown in the figure, and then build the string for each of them based on the method provided in Subsection~\ref{sec3_2}. After that, their hash values are calculated and shown in Table~\ref{table_hash_info}.
\newcommand{\tabincell}[2]{\begin{tabular}{@{}#1@{}}#2\end{tabular}}
\begin{table}[!htp]
	\renewcommand{\arraystretch}{1.2}
	\caption{The hash values of the two paths having the longest lengths in the first example.} \label{table_hash_info}
	\centering
	\begin{tabular}{|c|c|}
		\hline
		\multicolumn{1}{|c|}{\textbf{Path}} & \multicolumn{1}{c|}{\textbf{Hash Value}} \\
		\hline
		S-S2-S3-D1-D2-D3-F1-F2-F3-I-J-E
		& \tabincell{c}{79045a68420128099d0f13fc9d7cf\\1469ba0de0eb77f6525b01b1fa6} \\
		\hline
		S-S2-S3-B1-B2-B3-C1-C2-C3-I-J-E
		& \tabincell{c}{1a0fb84d6fb8142d9d6bf4396c37\\5e5075ba8e8f6cacaee6a759d043} \\
		\hline
	\end{tabular}
\end{table}

\begin{figure*}[!htp]
	\centering
	\includegraphics[width=2\columnwidth]{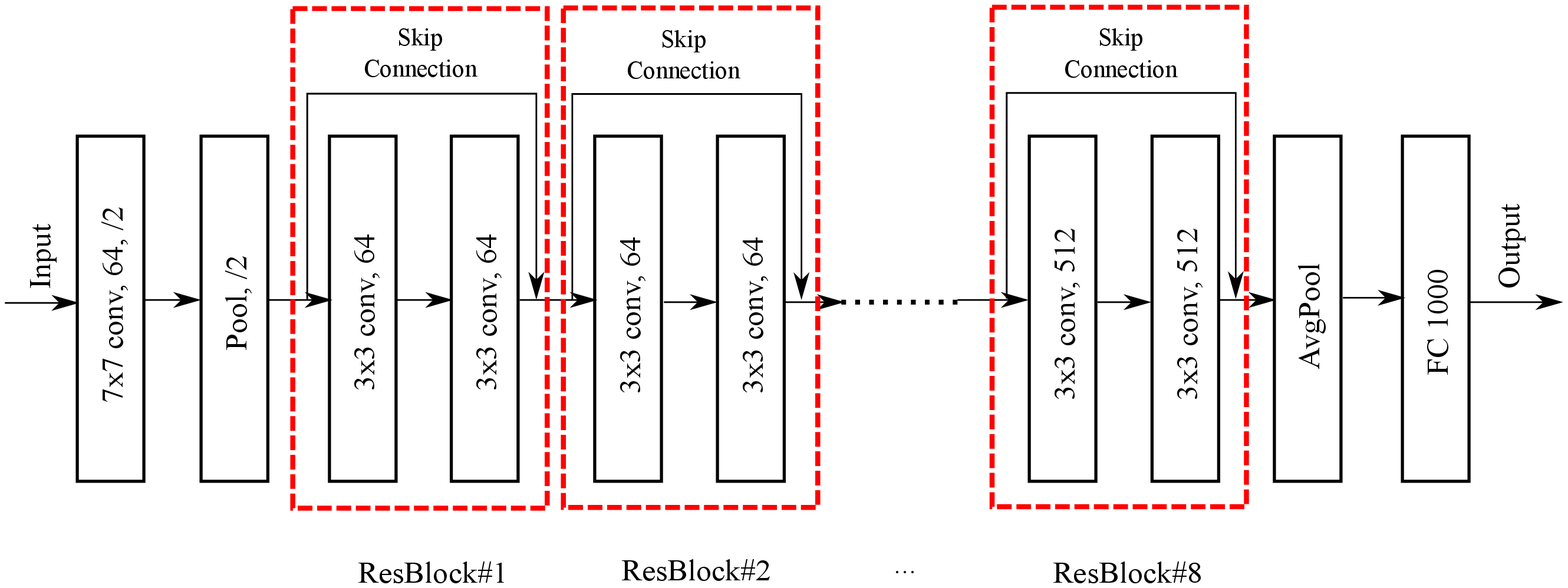}\\
	\caption{The image illustrating the architecture of ResNet18, including a convolutional layer, a max pooling layer, eight ResBlocks, an average pooling layer, and a fully-connected layer. }\label{fig_resnet}
\end{figure*}

\subsubsection{Numbering Layer Position} As can be seen from Table~\ref{table_hash_info}, the order of the two paths should be ``S-S2-S3-D1-D2-D3-F1-F2-F3-I-J-E'' and ``S-S2-S3-B1-B2-B3-C1-C2-C3-I-J-E'' based on the orders of their hash values. Consequently, the number of these layers are 1..25 for ``S'', ``S2'', ``S3'', ``D1'', ``D2'', ``D3'', ``F1'', ``F2'', ``F3'', ``I'', ``J'', ``B1'', ``B2'', ``B3'', ``C1'', ``C2'', ``C3'', ``G1'', ``G2'', ``G3'', ``H'', ``A1'', ``A2'', ``A3'', and ``E'', respectively. At this stage, all the information of each layer described by the corresponding ArcUnits are available, and the whole description of this CNN can be generated, and is shown in Table~\ref{table_text_example1}.

\begin{table*}[!htp]
	\renewcommand{\arraystretch}{1}	
	\caption{The description of the first example based on the proposed ArcText algorithm.} \label{table_text_example1}
	\centering
	\begin{tabular}{|p{1.8\columnwidth}|} 
		\hline
		id:1;in\_size:32-32-3;out\_size:32-32-3;kernel:1-1;stride:1-1;padding:0-0-0-0-0-0-0-0;dilation:1;groups:1;bias\_used:No;connect\_to:2\\
		id:2;name:ReLU;in\_size:32-32-3;out\_size:32-32-3;value:Null;connect\_to:3\\
		id:3;name:BN;in\_size:32-32-3;out\_size:32-32-3;value:Null;connect\_to:4-12-18-22\\
		id:4;in\_size:32-32-3;out\_size:16-16-3;kernel:2-2;stride:2-2;padding:0-0-0-0-0-0-0-0;dilation:1;groups:1;bias\_used:No;connect\_to:5\\
		id:5;name:ReLU;in\_size:16-16-3;out\_size:16-16-3;value:Null;connect\_to:6\\
		id:6;name:BN;in\_size:16-16-3;out\_size:16-16-3;value:Null;connect\_to:7\\
		id:7;in\_size:16-16-3;out\_size:16-16-10;kernel:1-1;stride:1-1;padding:0-0-0-0-0-0-0-0;dilation:1;groups:1;bias\_used:No;connect\_to:8\\
		id:8;name:ReLU;in\_size:16-16-10;out\_size:16-16-10;value:Null;connect\_to:9\\
		id:9;name:BN;in\_size:16-16-10;out\_size:16-16-10;value:Null;connect\_to:10\\
		id:10;name:Addition;in\_size:16-16-10;out\_size:16-16-10;value:Null;connect\_to:11\\
		id:11;in\_size:2560;out\_size:512;act\_fun:ReLU;connect\_to:25\\
		id:12;in\_size:32-32-3;out\_size:32-32-10;kernel:1-1;stride:1-1;padding:0-0-0-0-0-0-0-0;dilation:1;groups:1;bias\_used:No;connect\_to:13\\
		id:13;name:ReLU;in\_size:32-32-10;out\_size:32-32-10;value:Null;connect\_to:14\\
		id:14;name:BN;in\_size:32-32-10;out\_size:32-32-10;value:Null;connect\_to;15\\
		id:15;in\_size:32-32-10;out\_size:16-16-10;kernel:2-2;stride:2-2;padding:0-0-0-0-0-0-0-0;dilation:1;groups:1;bias\_used:No;connect\_to:16\\
		id:16;name:ReLU;in\_size:16-16-10;out\_size:16-16-10;value:Null;connect\_to:17\\
		id:17;name:BN;in\_size:16-16-10;out\_size:16-16-10;value:Null;connect\_to:10\\
		id:18;in\_size:32-32-3;out\_size:31-31-10;kernel:2-2;stride:1-1;padding:0-0-0-0-0-0-0-0;dilation:1;groups:1;bias\_used:No;connect\_to:19\\
		id:19;name:BN;in\_size:31-31-10;out\_size:31-31-10;value:Null;connect\_to:20\\
		id:20;name:ReLU;in\_size:31-31-10;out\_size:31-31-10;value:Null;connect\_to:21\\
		id:21;type:Max;in\_size:31-31-10;out\_size:16-16-10;kernel:2-2;stride:2-2;padding:1-0-1-0;dilation:1;bias\_used:No;connect\_to:10\\
		id:22;in\_size:32-32-3;out\_size:16-16-10;kernel:2-2;stride:2-2;padding:0-0-0-0-0-0-0-0;dilation:1;groups:1;bias\_used:No;connect\_to:23\\
		id:23;name:ReLU;in\_size:16-16-10;out\_size:16-16-10;value:Null;connect\_to:24\\
		id:24;name:BN;in\_size:16-16-10;out\_size:16-16-10;value:Null;connect\_to;10\\
		id:25;name:Dropout;in\_size:512;out\_size:512;value:0.5;connect\_to:Null
		
		\\	
		\hline
		
	\end{tabular}
\end{table*}

\subsection{Example for the Second Representative}
It has been widely recognized that the promising performance of ResNet is mainly contributed by its skip connections~\cite{he2016deep,huang2017densely,sun2017evolving}. Specifically, the connections in the traditional CNNs only occur at the neurons at the adjacent layers, while the skip connections provide the fact that the connections bridge the neurons from the layers which are not adjacent. In addition, the traditional connections often are connected with the weights equal to one, while the skip connections have the weights equal to one. This is also the main reason that the skip connections are called as the ``identity connections''. Many researchers have put concerns on investigating the mechanism of the success of such kinds of connections, while there is still no clear and unified conclusions yet. Commonly, a large proportion of researchers believe that the skip connections could address the gradient vanishing problem, thus, the resulted CNNs could enhance the performance~\cite{gers1999learning,hochreiter1997long}. Interested readers could refer to \cite{gers1999learning} for more details about the gradient vanishing problem.

\begin{figure}[!htp]
	\centering
	\includegraphics[width=0.6\columnwidth]{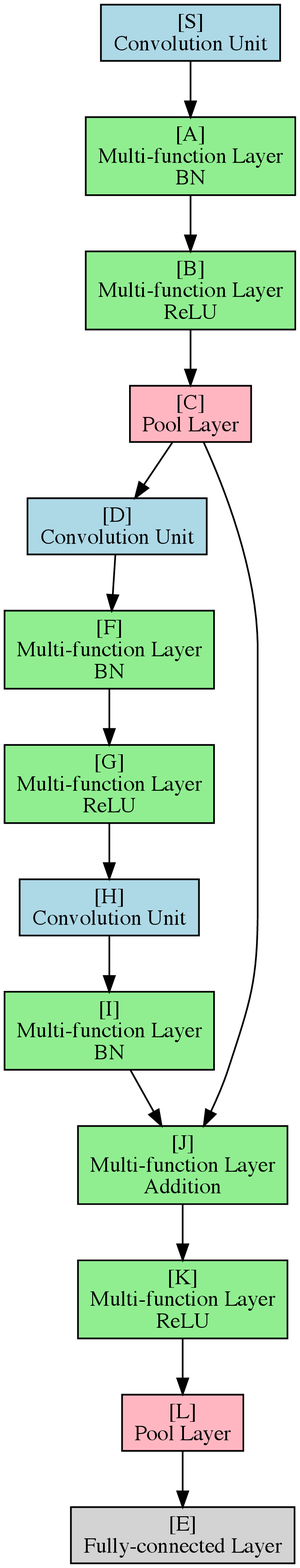}\\
	\caption{The architecture information of the second example CNN (ResNet4). In this figure, each rectangle denotes a node of the CNN. The word inside each block shows the name and the information of the node.}\label{fig_resnetblock}
\end{figure}

The building block of ResNet refers to the minimal unit containing a skip connection, which is also called as the ResBlock. There have been multiple variants of ResNet, among which they differ in the numbers of ResBlocks. The main reason is that the large number of ResBlock, the higher computational complexity will be. In some situations, the problem at hand may not need a larger number of ResBlocks. Fig.~\ref{fig_resnet} shows a variant of ResNet which contains eight ResBlocks. This variant is called as the ResNet18 because it has a total of 18 layers having learnable weights. Specifically, there are two layers before the first ResBlock and another two layers after the last ResBlock. For each ResBlock, there are also two layers. because the pooling layers have no learnable weights, this variant has ($1+8\times 2 + 1=18$) layers.  In this figure, the numbers in rectangles (in addition to the first and the last ones) refer to the configuration of the corresponding layers. Particularly, the rectangle marked as ''FC 1000'' refers to a fully-connected layer having 1000 neurons, and the "AvgPool" means an average pooling layer. For the other rectangles, their layer types can be reflected by the work inside them. For example, the second rectangle refers to a convolutional layer which has $64$ kernels and each kernel is with the size of $7\times 7$. The symbol of ``$\backslash2$'' means the stride is $2\times 2$. If the rectangles have no such symbols, it means that the corresponding layer adopts the stride of $1\times 1$. In this example, we will use the proposed ArcText to describe the whole architecture of ResNet18 in addition to the last five blocks for the reason of simplicity. Following the principle of naming the variants of ResNet, we name this example as ResNet4 and also use this name for the following presentation. Particularly, the architecture of ResNet4 is shown in Fig.~\ref{fig_resnetblock}, and its configurations are provided in Tables~\ref{table_conv_example2} to \ref{table_mf_example2}. Please note that the padding information listed in Tables~\ref{table_conv_example2} and \ref{table_pool_example2_p2} denote that the padding configurations at the four directions share the same information.

\subsubsection{Constructing Graph}
As have highlighted that the example will be elaborated on ResNet4, thus, the graph is also constructed on the ResNet4 model. The construction is also straightforward, and can be viewed from Fig.~\ref{fig_resnetblock}, where the input layer and the output layer have already named as ``S'' and ``E'', respectively, based on the details provided in Algorithm~\ref{alg_position_location}. In addition, the words inside each rectangles of Fig.~\ref{fig_resnetblock} have the same meaning as those in Fig.~\ref{fig_example} which have been detailed in Subsection~\ref{sec_first_example}.

\begin{table}[htp]
	\renewcommand{\arraystretch}{1.2}
	\caption{The information of the convolutional layers of the second example (Part I).} \label{table_conv_example2}
	\centering
	\begin{tabular}{|c|c|c|c|c|c|} 
		\hline
		Id&in\_size&out\_size&kernel&stride&padding\\
		\hline
		S&224,224,3&112,112,64&7,7&2,2&0,3\\
		\hline
		D&56,56,64&56,56,64&3,3&1,1&0,1\\
		\hline
		H&56,56,64&56,56,64&3,3&1,1&0,1\\
		\hline
	\end{tabular}
\end{table}

\begin{table}[htp]
	\renewcommand{\arraystretch}{1.2}
	\caption{The information of the convolutional layers of the second example (Part II).} \label{table_conv_example2_p2}
	\centering
	\begin{tabular}{|c|c|c|c|c|} 
		\hline
		Id&dilation&groups&bias\_used&connect\_to \\
		\hline
		S&1&1&No&A\\
		\hline
		D&1&1&No&F\\
		\hline
		H&1&1&No&I\\
		\hline
	\end{tabular}
\end{table}

\begin{table}[!htp]
	\renewcommand{\arraystretch}{1.2}
	\caption{The information of the pooling layers of the second example (Part I).} \label{table_pool_example2}
	\centering
	\begin{tabular}{|c|c|c|c|c|c|} 
		\hline
		Id&type&in\_size&out\_size&kernel&stride \\
		\hline
		C&Max&112,112,64&56,56,64&3,3&2,2\\
		\hline
		L&Avg&56,56,64&1,1,64&56,56&1,1\\
		\hline
	\end{tabular}
\end{table}

\begin{table}[!htp]
	\renewcommand{\arraystretch}{1.2}
	\caption{The information of the pooling layers of the second example (Part II).} \label{table_pool_example2_p2}
	\centering
	\begin{tabular}{|c|c|c|c|c|} 
		\hline
		Id&padding&dilation&bias\_used&connect\_to \\
		\hline
		C&0,1&1&No&D,J\\
		\hline
		L&0,0&1&No&E\\
		\hline
	\end{tabular}
\end{table}

\begin{table}[!htp]
	\renewcommand{\arraystretch}{1.2}	
	\caption{The information of the fully-connected layers of the second example.} \label{table_full_example2}
	\centering
	\begin{tabular}{|c|c|c|c|c|} 
		\hline
		Id&in\_size&out\_size&act\_fun&connect\_to\\
		\hline
		E&64&1000&ReLU&Null\\
		\hline
	\end{tabular}
\end{table}

\begin{table}[!htp]
	\renewcommand{\arraystretch}{1.2}	
	\caption{The information of the other nodes of the second example.} \label{table_mf_example2}
	\centering
	\begin{tabular}{|c|c|c|c|c|c|} 
		\hline
		Id&Name&in\_size&out\_size&value&connect\_to\\
		\hline
		A&BN&112,112,64&112,112,64&Null&B\\
		\hline
		B&ReLU&112,112,64&112,112,64&Null&C\\
		\hline
		F&BN&56,56,64&56,56,64&Null&G\\
		\hline
		G&ReLU&56,56,64&56,56,64&Null&H\\
		\hline
		I&BN&56,56,64&56,56,64&Null&J\\
		\hline
		J&Addition&56,56,64&56,56,64&Null&K\\
		\hline
		K&ReLU&56,56,64&56,56,64&Null&L\\
		\hline
	\end{tabular}
\end{table}

\subsubsection{Finding Path}
As can bee see from Fig.~\ref{fig_resnetblock}, there is only one longest path in the graph which is ``S-A-B-C-D-F-G-H-I-J-K-L-E''. This also can be justified that the ResNet architecture is much simpler than the first example provided in Subsection~\ref{sec_first_example}. Because no paths have the same lengths in this example, we do not need to compute the hash values for the ordering.

\subsubsection{Numbering Layer Position}
Since there is only one longest path in Fig.~\ref{fig_resnetblock}, the number of these layer are 1..13 for ``S'', ``A'', ``B'', ``C'', ``D'', ``F'', ``G'', ``H'', ``I'', ``J'', ``K'', ``L'', ``E'', respectively. At this stage, the resulted description of ResNet4 can be easily obtained, which is shown in Table~\ref{table_text_example2}.

\begin{table*}[!htp]
	\renewcommand{\arraystretch}{1}	
	\caption{The description of the second example (ResNet4) based on the proposed ArcText algorithm.} \label{table_text_example2}
	\centering
	\begin{tabular}{|p{1.8\columnwidth}|} 
		\hline
		id:1;in\_size:224-224-3;out\_size:112-112-64;kernel:7-7;stride:2-2;padding:0-3-0-3-0-3-0-3;dilation:1;groups:1;bias\_used:No;connect\_to:2\\
		id:2;name:BN;in\_size:112-112-64;out\_size:112-112-64;value:Null;connect\_to:3\\
		id:3;name:ReLU;in\_size:112-112-64;out\_size:112-112-64;value:Null;connect\_to:4\\
		id:4;type:Max;in\_size:112-112-64;out\_size:56-56-64;kernel:3-3;stride:2-2;padding:1-1-1-1;dilation:1;bias\_used:No;connect\_to:5-10\\
		id:5;in\_size:56-56-64;out\_size:56-56-64;kernel:3-3;stride:1-1;padding:0-1-0-1-0-1-0-1;dilation:1;groups:1;bias\_used:No;connect\_to:6\\
		id:6;name:BN;in\_size:56-56-64;out\_size:56-56-64;value:Null;connect\_to:7\\
		id:7;name:ReLU;in\_size:56-56-64;out\_size:56-56-64;value:Null;connect\_to:8\\
		id:8;in\_size:56-56-64;out\_size:56-56-64;kernel:3-3;stride:1-1;padding:0-1-0-1-0-1-0-1;dilation:1;groups:1;bias\_used:No;connect\_to:9\\
		id:9;name:BN;in\_size:56-56-64;out\_size:56-56-64;value:Null;connect\_to:10\\
		id:10;name:Addition;in\_size:56-56-64;out\_size:56-56-64;value:Null;connect\_to:11\\
		id:11;name:ReLU;in\_size:56-56-64;out\_size:56-56-64;value:Null;connect\_to:12\\
		id:12;type:Avg;in\_size:56-56-64;out\_size:1-1-64;kernel:56-56;stride:1-1;padding:0-0-0-0;dilation:1;bias\_used:No;connect\_to:13\\
		id:13;in\_size:64;out\_size:1000;act\_fun:ReLU;connect\_to:Null
		\\	
		\hline
		
	\end{tabular}
\end{table*}

\section{Conclusion and Future Work}
\label{section_conclusion}
The goal of this paper is to propose a unified method of describing CNN architectures, enabling abundant CNN architectures available to be vectorized and then applied by various data mining algorithms, which can further promote the research on CNNs, such as discovering useful patterns of the deep architectures to significantly relieve the human expertise in manually designing CNN architectures, and finding the relationship between CNN architectures and their performance to address the computationally expensive problem of existing NAS algorithms. The goal has been achieved by the proposed ArcText method. Specifically, four units have been designed in ArcText, to describe the detailed information of the nodes in CNNs. In addition, a novel component has also been developed to assign a unique order of nodes in the CNN, ensuring the constant topology information obtained whenever the CNN is described. Furthermore, examples have been provided in this paper to illustrate how ArcText works given the popular CNNs. The proposed ArcText method can be viewed as a grammar rule of CNN architecture description based on language, thus the advanced natural language processing techniques can be easily built upon the proposed algorithm to design advanced applications. Moreover, newly generated CNN architectures can be easily shared and exchanged via the proposed method to a public repository, providing sufficient data for data mining algorithms. The construction of the repository will be reserved as our future endeavor.


\end{document}